\documentclass[10pt,twocolumn,twoside,letterpaper]{IEEEtran}

\usepackage{geometry}
\usepackage{amsmath}
\usepackage{url}
\usepackage{caption}
\usepackage{lineno}
\makeatletter
\def\ps@IEEEtitlepagestyle{
  \def\@oddfoot{\mycopyrightnotice}
  \def\@evenfoot{}
}
\def\mycopyrightnotice{
}

\usepackage{amssymb}
\usepackage{url}
\usepackage{siunitx}
\usepackage{graphicx}
\usepackage{subfig}
\RequirePackage[colorlinks=true
,urlcolor=blue
,anchorcolor=blue
,citecolor=blue
,filecolor=blue
,linkcolor=blue
,menucolor=blue
,pagecolor=blue
,linktocpage=true
,pdfa=true
]{hyperref}
\begin{document}
\title{Exploring Advanced Detector Technologies for Muon Radiography Applications
}
%
%
%

\author{A. Samalan,~\IEEEmembership{Member,~IEEE}, Y. Assran, C. A. Diaz Escorcia, B. EIMahdy, Y. Hong,~\IEEEmembership{Member,~IEEE}, G. Prithivraj, C. Rendon, D. Samuel, M. Tytgat,~\IEEEmembership{Member,~IEEE}
\thanks{Manuscript received December 9, 2022. This work was supported in part by the FWO (Belgium).}
\thanks{A. Samalan, C. A. Diaz Escorcia,  Y. Hong,  C. Rendon and M. Tytgat are with Department of Physics and Astronomy, Ghent University, Gent-9000,
Belgium (e-mail: amrutha.samalan@ugent.be, carlosandres.diaz@ugent.be, yanwen.hong@ugent.be,  cesar.rendon@ugent.be, michael.tytgat@ugent.be).}
\thanks{B. EIMahdy is with Centre for Theoretical Physics, The British University in Egypt, El Sherouk City, Cairo-11837, Egypt and Department of physics, Faculty of Science, Helwan University, Ain Helwan (University Campus), Caoiro-11795, Egypt (e-mail: basma.mahdy@bue.edu.eg)}
\thanks{Y. Assran is with Centre for Theoretical Physics, The British University in Egypt, El Sherouk City, Cairo-11837, Egypt and Faculty of Petroleum and Mining Engineering, Suez University, Suez, Governorate-8151650, Egypt(e-mail: Yasser.Assran@cern.ch)}
\thanks{G. Prithivraj and D. Samuel are with the Department of
Physics, School of Physical Sciences, Central University of Karnataka, Aland
Road, Gulbarga Dist- 585367, India (e-mail: prithivrajastro@gmail.com, deepaksamuel@cuk.ac.in).}
}

\maketitle

\pagenumbering{gobble}

\begin{abstract}
Muon radiography often referred to as muography,
is an imaging technique that uses freely available cosmic-ray
muons to study the interior structure of natural or man-made
large-scale objects. The amount of multidisciplinary applications
of this technique keeps increasing over time and a variety of basic
detector types have already been used in the construction of
muon telescopes. Here, we are investigating the use of advanced
gaseous detectors for muography. As our basic solution, given its
robustness and ease of operation in remote, outdoor
environments, a scintillator-based muon telescope with silicon
photomultiplier readout is being developed. To enhance the
telescope performance, we are proposing the use of Multi-gap
Resistive Plate Chambers (mRPCs) and Thick Gas Electron
Multipliers (THGEMs). While the former offer superior time
resolution which could be beneficial for detector background
rejection, the latter detector type offers excellent spatial
resolution, can be manufactured at low cost and operated with a
simple gas mixture. Currently, prototype detector planes for each
of these proposed types are being designed and constructed, and
initial performance tests are in progress. In parallel, a Geant4-
based muon telescope simulation is being developed, which will
enable us to e.g. optimize our telescope geometry and study the
use of superior time resolution for background rejection. The
design and status of the three detector prototype planes and the
muon telescope, along with the initial results of their performance
tests and of the Geant4 simulation studies are reported.
\end{abstract}


\section{Introduction}
%
%
%
%
\IEEEPARstart{C}{osmic}-ray muons are generated when high energy primary
cosmic rays interact with our earth’s atmosphere. Due to the
relative hardness of the atmospheric muon spectrum, most of
these muons reach sea level. Profiting from their long relativistic 
lifetime and their high penetration capability, they can be exploited as 
imaging probes of the internal structure of large-scale objects~\cite{bonomi}. 
A standard 2D muographic image represents an average density profile of the target along the line of sight of the muons and can be derived from measurements of the transmitted 
muon flux through the object; the average density $\rho$ of the
target material is then calculated from the opacity $\chi$ using
$\chi$  = x$\rho$, where $x$ is the length of rock that the muon traversed. Using multiple viewpoints in principle allows to create 3D images of the object.  
Muographic imaging is increasingly being applied in many areas including for example volcanology, archaeology, civil engineering, industry, mining, nuclear waste surveys, homeland
security, natural hazard monitoring and several more. 

Muon telescopes in muography are traditionally built using technology based on either nuclear emulsions, scintillators or gaseous detectors (see e.g.~\cite{procureur2018}). Each of these types comes with its own set of features and advantages, and one type may be preferred over the other depending on the specific application and the target object and its environment. 
Here, to exploit the particular advantages offered by certain technologies we aim to employ multiple detector types together in one single, highly-performing muon telescope.

Currently, three different kinds of muon detectors are being considered.  
As our basic solution, a 
scintillator-based muon telescope made of plastic scintillator bars
coupled to Multi-Pixel Photon Counters (MPPCs) is under development. Such a system is relatively low cost and low power consumption and offers robustness and sustainability under
harsh environments. It appears to
be the solution that has so far been most commonly used in
muography setups.
To improve upon this basic system in various areas we are considering the usage of advanced gaseous detectors.
Solutions based on e.g. Multi-wire Proportional Chambers, Resistive Plate Chambers, Micromegas already exist or are being developed (see e.g.~\cite{varga},~\cite{muongraphyapplication},~\cite{portablerpc}). Here, we are considering Multi-gap Resistive Plate Chambers (mRPCs), and Thick Gas Electron
Multipliers (THGEMs) with the specific main aim to simultaneously improve
the muon telescope performance in terms of time and spatial
resolution. mRPCs may play a crucial role in detector background rejection (detector noise, backwards-going muons ...)
due to their ability to provide sub-ns timing information.
THGEMs on the other hand offer an excellent position resolution in the ~100$\mu$m range which will enhance the muographic
image resolution. Compared to standard GEMs their design is
more robust and allows manufacturing in an economical way.
Compared to RPCs, they can be operated using simple gas
mixtures or even mono-gases, which greatly simplifies their operation in
outdoor environments.

At Ghent University, prototype detector planes using each of the three aforementioned technologies are being designed, constructed and
their performance is being evaluated in the lab using cosmic-ray muons and/or X-rays. In addition, a Geant4-based~\cite{geant4} simulation of a generic 4-plane muon telescope is being developed
to fine-tune the detector and telescope geometry and to study the effectiveness
of using improved time information in the background rejection. Finally, a Garfield++-based~\cite{garfield} simulation of THGEMs is being used to study and optimize the detector performance in terms of e.g. hole dimensions, detector gap sizes, field values, gas mixtures. 
Section II describes the details
of the scintillator system, while the details of the mRPC and
THGEM prototypes are described in sections III and IV respectively.

\section{Scintillator detector}
\begin{figure}[htpb!]%
    \centering
    \subfloat[\centering ]{{\includegraphics[width=1.7in]{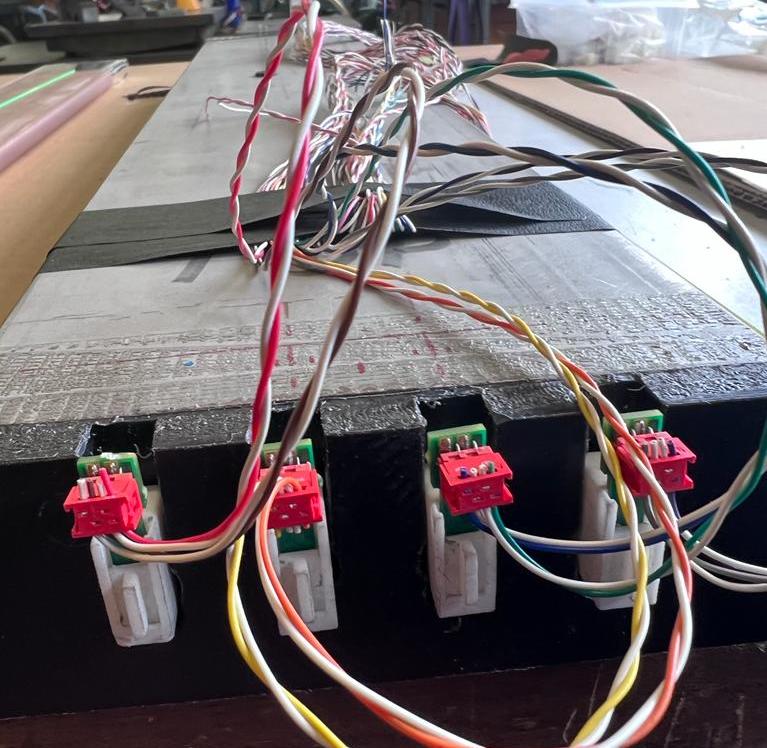} }}%
    \qquad
    \subfloat[\centering ]{{\includegraphics[width=1.3in]{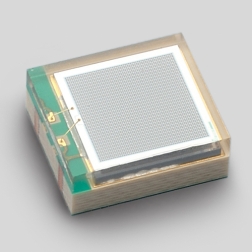} }}%
    \caption{(a) Scintillator prototype detector equipped with four MPPCs. (b) Hamamatsu MPPC with photosensitive area of $3\times 3$~mm$^2$.}
    \label{fig:scint}%
\end{figure}

Our present scintillator prototype is made with 
rectangular plastic scintillator bars with an active area of $56\times 10\times 1.4$~cm$^3$. Bicron BC-620 diffuse reflector paint is applied on one edge of the bars to guide the light in one direction. 
Scintillation photons emitted inside each bar are
collected through four BCF-91A wavelength shifting fibers of type BCF-91A~\cite{wlsfibers} embedded in equidistant grooves in the scintillator surface. Each fiber is directly coupled to a Hamamatsu S12572-010~\cite{SiPM}
Multi-Pixel Photon Counter (MPPC) with a  photosensitive area of $3\times 3$~mm$^2$. The whole setup is wrapped in copper one-way mirror film and then placed inside an aluminium box for noise reduction. 
Figure~\ref{fig:scint} shows the scintillator prototype equipped with four MPPCs and the photosensitive area of the MPPC.
MPPCs utilise tiny pixels arrayed at high
densities which results in a low recovery time and a wide
dynamic range. 
Compared to more traditional photomultipliers, these MPPCs are available at a much lower cost, they are less power consuming ($\sim$10 times less) and are very compact in terms of space for installation.

\begin{figure}[htpb!]
\centering
\includegraphics[width=2.5in]{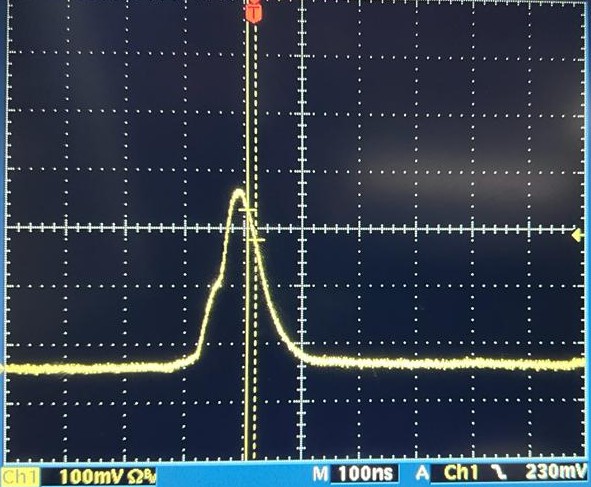}
\caption{The raw signal obtained from the Hamamatsu S12572-010 MPPC using an amplifier plus a peak detector circuit.}
\label{fig:signal}
\end{figure}

\begin{figure}[htpb!]
\centering
\includegraphics[width=3.3in]{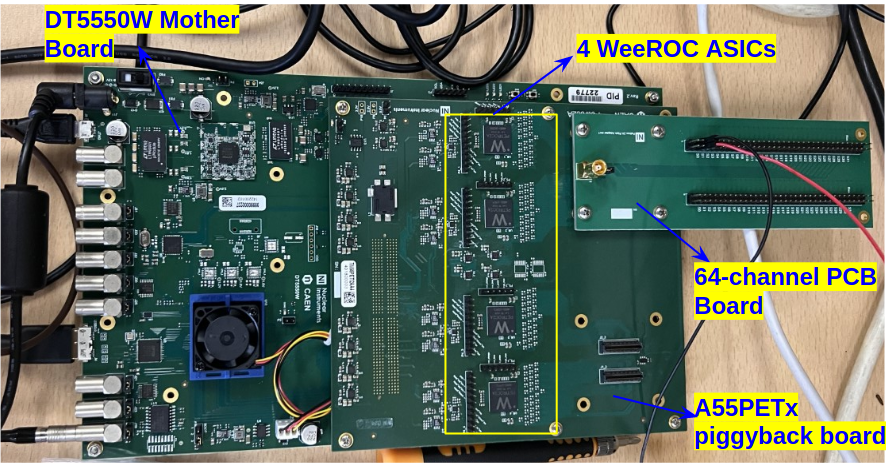}
\caption{The DAQ setup for the scintillator detector. The DT5550W motherboard, the A55PET4 piggyback board hosting four PETIROC2A WeeROC ASICs and  64-channel PCB board plugged on to the piggyback board are shown.}
\label{fig:daq}
\end{figure}

The MPPC in our setup has a breakdown voltage V$_{BR}$= 65$\pm$10~V. When a muon-induced photo-avalanche occurs, a positive MPPC pulse is created of width $\sim$500~ns and height 20-80~mV. To study the raw output signal, the pulse is sent
to a non-inverting LT2601 amplifier, where it gets amplified by
approximately a factor of six and then passed to a peak
detector circuit that
outputs a pulse that rises to the peak of the amplified pulse but decays slowly over a period
of roughly 100~$\mu s$. Figure~\ref{fig:signal} shows the MPPC signal obtained using the peak detection circuit.
Eventually, custom made muon-boards capable of hosting 64 MPPCs interfaced to a CAEN-based data acquisition (DAQ) system will be used for the final signal extraction and processing. 
The DAQ system is currently in the calibration stage~\cite{Kouzes}. It consists of a CAEN DT5550W motherboard hosting a programmable FPGA and a CAEN A55PET4 piggyback board with four PETIROC2A WeeROC ASICs~\cite{CAENDAQ}. The system  can perform energy and high-resolution time measurements with SiPMs. The DT5550W can be used either as an evaluation system for WeeROC ASICs or as a full-featured readout system for SiPMs. The readout of the PETIROC ASIC is entirely managed by the firmware preloaded on
the DT5550W FPGA. The piggyback board can be easily plugged onto the motherboard to provide a complete DAQ for up to 128 MPPCs. Along with the piggyback board, there are two 64-channel Printed Circuit Boards (PCBs) which allows the connection of a single MPPC or an array of MPPCs to the board. The PCB provides 64 signal readout lines (Sx) and 64 sensor bias (B) lines. Each bias line is filtered with a 1~k$\Omega$ resistor and a 100~nF capacitor for the MPPC connection. This PCB is designed to be pressure-plugged onto the input connectors of the piggyback board. The custom made muon-board will be connected to the DAQ through these PCB boards. Figure~\ref{fig:daq} shows the DAQ setup and one of the PCB boards interfaced with it.

The major tasks currently ongoing with this prototype are the calibration of the MPPCs, quantification of
the noise levels, and optimisation of optical shielding
required to improve the position resolution. Ultimately, we aim to develop a modular a scintillator-based telescope with four orthogonally placed X-Y planes
of active area around $1\times1 $~m. Although such a system is suitable for many muography applications, 
advanced gaseous detectors are being developed in parallel to complement the scintillator planes for the following main reasons:
\begin{itemize}
\item MPPCs have a highly temperature-dependent
breakdown voltage which consequently affects the gain and
dark count rate. This leads to the necessity of a
temperature monitoring and stabilization system and/or a continuous readjustment of the
operating voltage.
\item The scintillator solution offers a relatively low position, angular and time resolution compared to what is achievable with gaseous detectors.
\end{itemize}

\section{Multi-gap Resistive Plate Chambers}
Resistive Plate Chambers (RPCs) are gaseous detectors that are widely used for muon
detection. They are operated at a high electric field
and are mainly used for accurate timing and fast triggering due
to their excellent intrinsic time resolution. Compared to
standard RPCs, mRPCs have an advanced layout with a wide
gas gap subdivided into multiple very thin gaps of order $\sim 150-300$~$\mu$m using intermediate thin glass plates in between the main electrodes, which results in
an improved time resolution in the range of 25-50~ps and high
rate capability~\cite{mRPC}. 
A first mRPC prototype chamber is currently under development. We are targeting a detector with a 2D position readout reaching a resolution well below 1~mm, in combination with a time resolution below 100~ps. Our chamber is designed with a low-volume, leak tight gas compartment, which in principle allows the detector to operate in sealed, zero-flow mode.  
In the context of an independent muography project~\cite{rmid} we are also developing a double-gap version of our detector with a similar advanced design as discussed here, where this chamber will be used to benchmark the performance of our mRPC.

\begin{figure}[htpb!]
\centering
\includegraphics[width=3.5in]{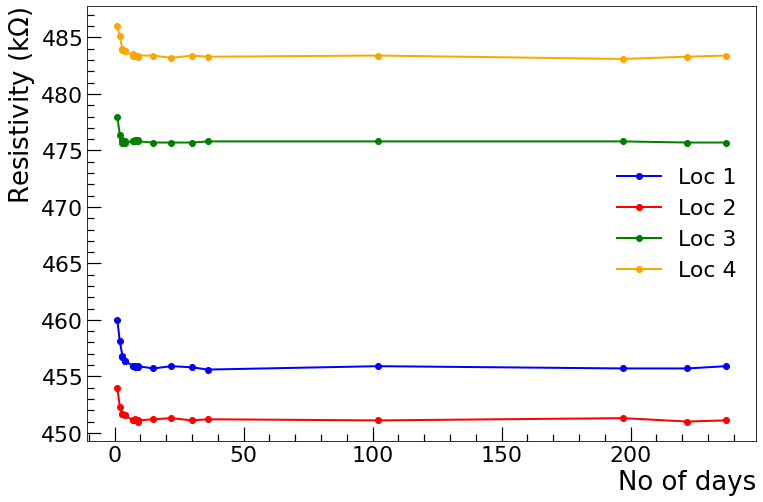}
\caption{Results of the resistivity monitored for a single resistive plate from September 2021 to May 2022. While a small drop was observed during the initial days after painting, it is seen to remain stable afterwards. Each line in the graph corresponds to a different measurement location on the glass plate.}
\label{fig:resmeas}
\end{figure}

\begin{figure}[htpb!]
  \centering
  \subfloat[a][]{\includegraphics[width=2.8in]{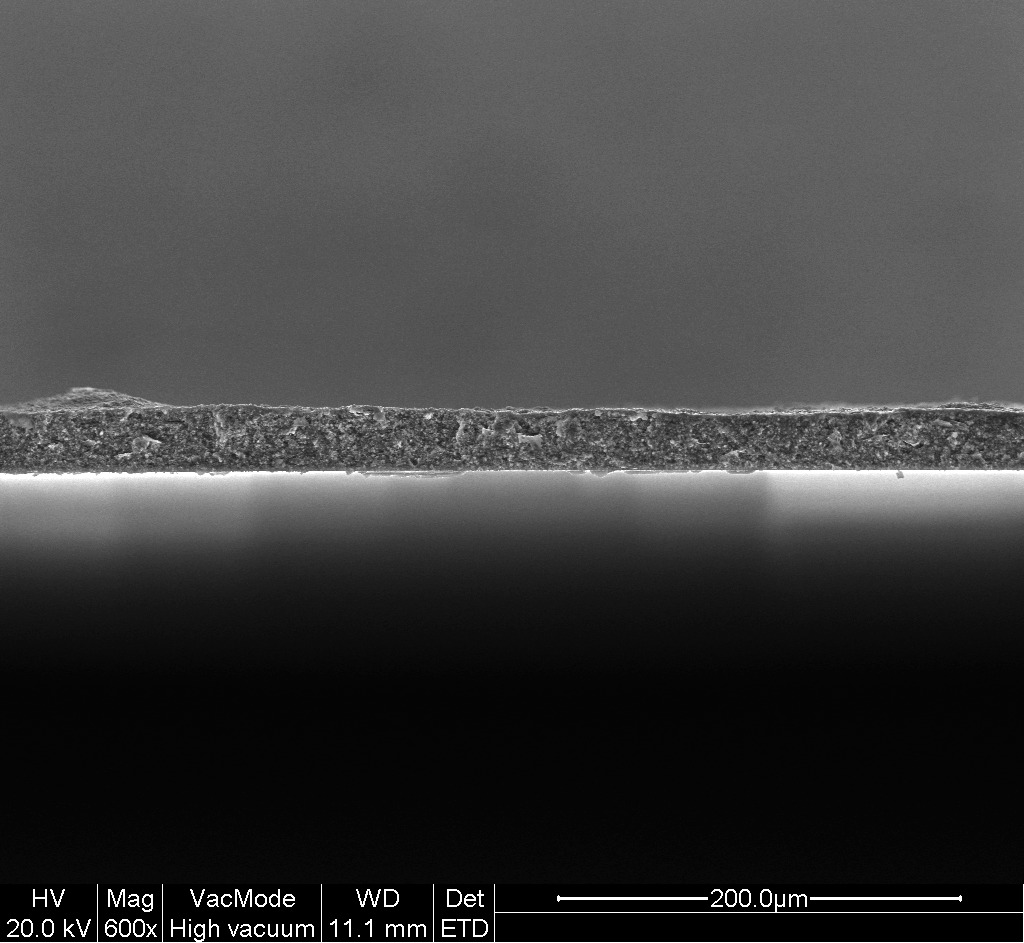} \label{fig:a}} \\
  \subfloat[b][]{\includegraphics[width=2.8in]{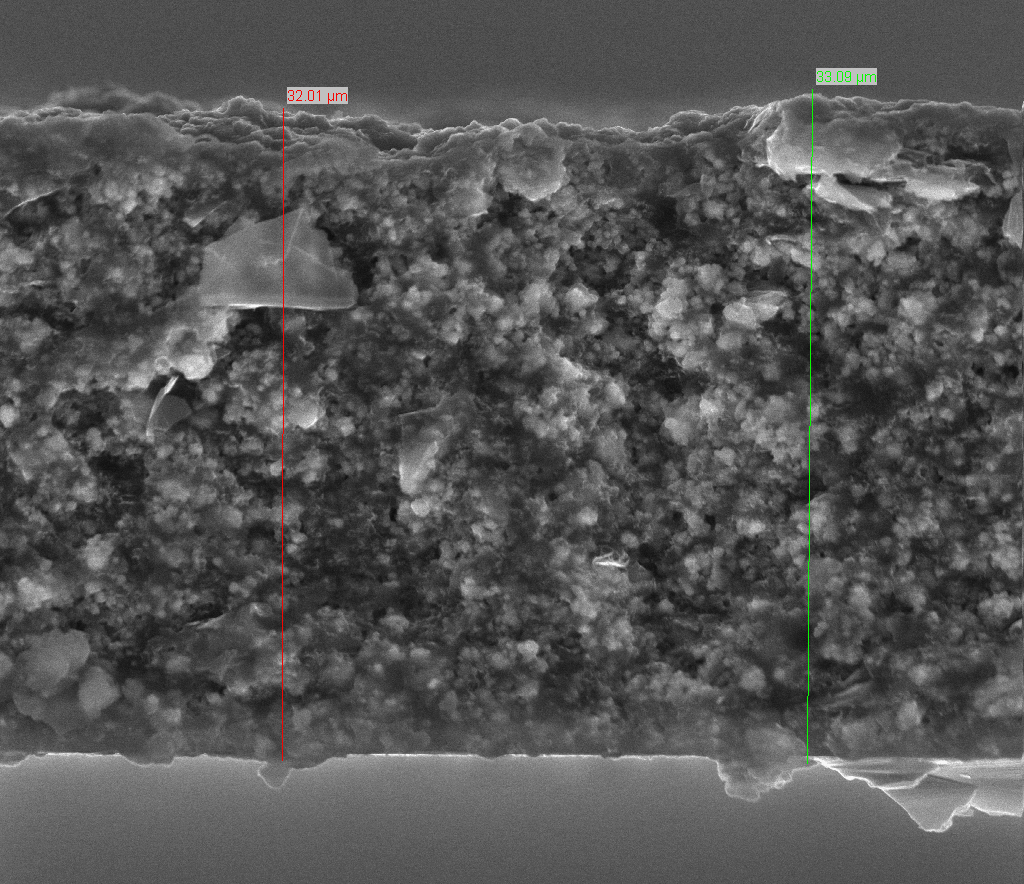} \label{fig:b}}
  \caption{(a). The SEM image of the resistive layer coating for the thickness measurement. (b) SEM image of a small region of the resistive coating ($\sim 45$~$\mu m$) for fine thickness measurement.} \label{fig:sem}
\end{figure}

For our mRPC prototype, the floating glass based resistive electrodes with active area of 
$28\times 28$~cm$^2$ are coated on one side with a two-component paint layer using an in-house technique. For this resistive coating, a mixture of Electrodag 6017SS conductive and Electrodag PM-404 resistive paint is used ~\cite{paint}. The paint mixture is sprayed onto the glass plates via a hand-held painting gun operated under a nearly constant pressure of 4.5 to 5~bar supplied by an air compressor. After painting, the glass plates are immediately baked in an oven for 20 minutes at 100$^{\circ}$C to allow the coating to dry. A uniform resistive layer
in terms of thickness as well as resistivity is achieved, which
is up to the quality level of industrially developed resistive
plates. The surface resistivity of the plates varies from 0.4 to 1~M$\Omega /$$\Box$ for mass ratios ($r_c$) of the applied paint mixture ranging from 0.86 to 0.89. $r_c$ is defined as the ratio of the mass of the conductive paint ($m_c$) to the total mass of the mixture:
\begin{equation}
\label{eqn1}
r_c =\frac{m_c}{m_r+m_c}
\end{equation}
The surface resistivity of the coated plates is under continuous
monitoring and is so far seen to remain stable. Figure~\ref{fig:resmeas} shows the resistivity measurements performed from September 2021 to May 2022. The resistivity was measured at four different locations across a single glass plate. 
The measured resistivity values across a single plate are uniform within less than 10~\%.
Since our glass coating procedure is a manual one, the thickness of the layer was measured using a Scanning Electron Microscope (SEM) to study its uniformity. Figure~\ref{fig:a} shows an image of the resistive layer obtained using the SEM method and Figure~\ref{fig:b} shows a zoomed view of a small part of the coating ($\sim 45$~$\mu m$ wide). The SEM measurements confirmed that the current coating technique effectively results in a uniform resistive layer with a thickness varying from 30 to 35~$\mu m$.

\begin{figure}[htpb!]
\centering
\includegraphics[width=3.5in]{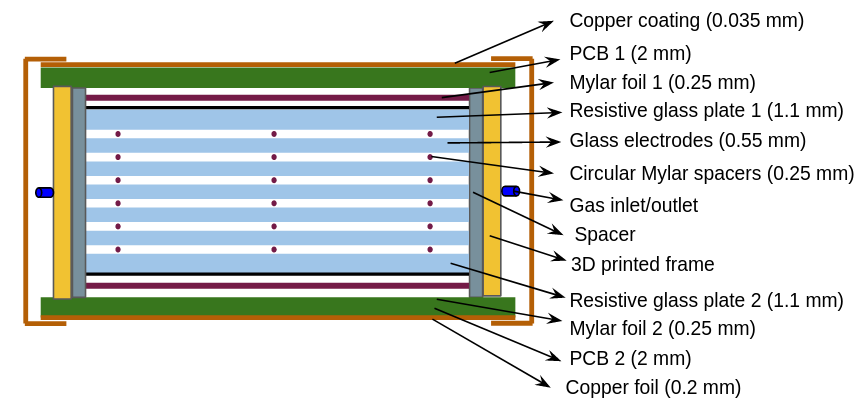}
\caption{Cross-sectional view of the mRPC prototype detector.}
\label{mrpc_sketch}
\end{figure}

Our mRPC prototype layout consists of a stack of 7 glass plates, i.e. six gas gaps. The top and bottom glass electrode plates have a thickness of 1.2~mm, while the intermediate plates in between those electrodes are floating glass plates of thickness 0.5~mm. The gas gaps will be constructed with circular pieces of mylar foil of thickness 200~$\mu m$ and a diameter of 5~mm. The whole glass stack will be placed inside a rectangular gas tight detector frame. To extract 2D positional information, the detector frame including the glass stack will be sandwiched two orthogonally oriented readout Printed Circuit Boards (PCBs) with, in a first test version, 32 copper strips of 0.8 mm pitch.
0.25~$\mu m$ thick Mylar foils will be placed between the PCBs and the glass electrodes for insulation. The back (outer) side of each PCB is coated with Copper, such that the top and bottom PCBs can be interconnected using Copper foils on all four sides of the detector. This arrangement will serve as a Faraday cage for noise reduction. Figure~\ref{mrpc_sketch} shows the cross-sectional view of the mRPC prototype.

\begin{figure}[htpb!]
\centering
\includegraphics[width=3in]{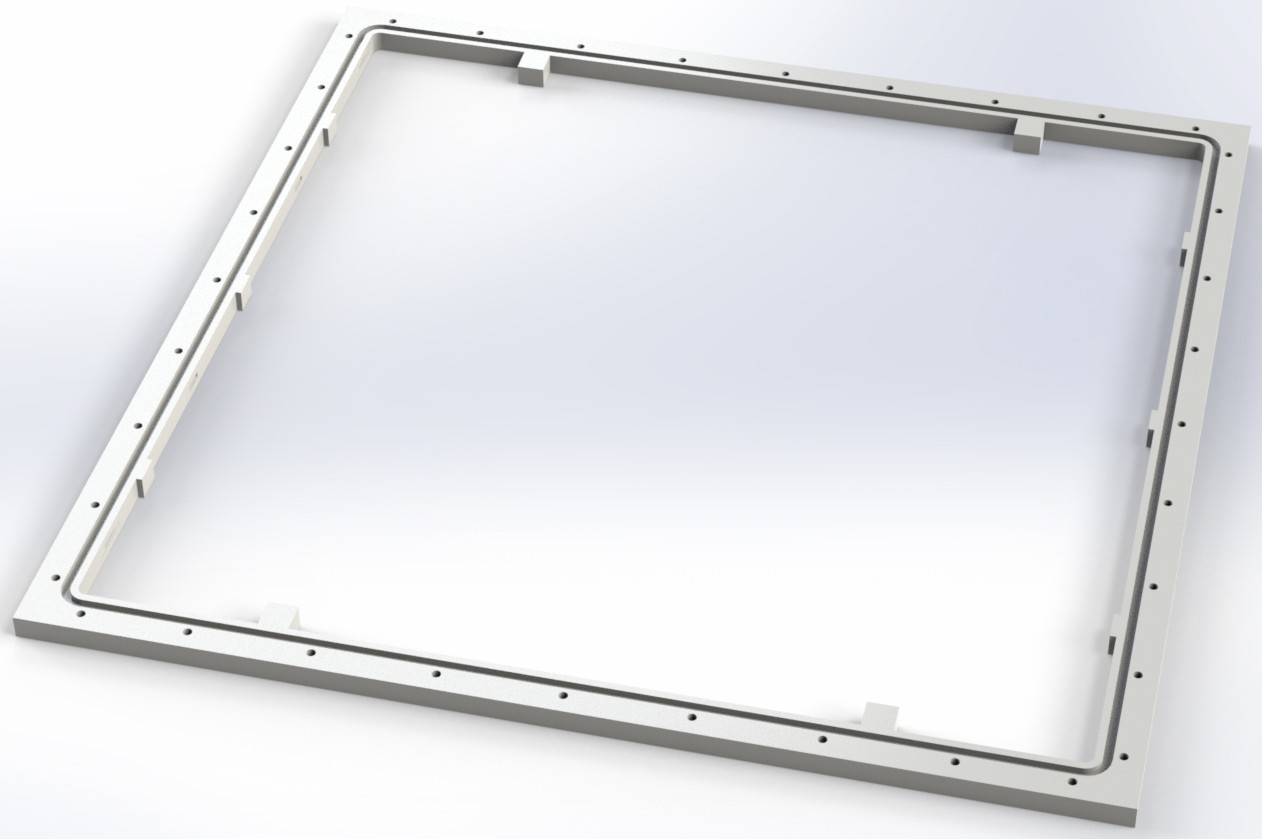}
\caption{The 3D CAD model of the mRPC detector frame.}
\label{fig:frame}
\end{figure}

\begin{figure}[htpb!]
\centering
\includegraphics[width=3.5in]{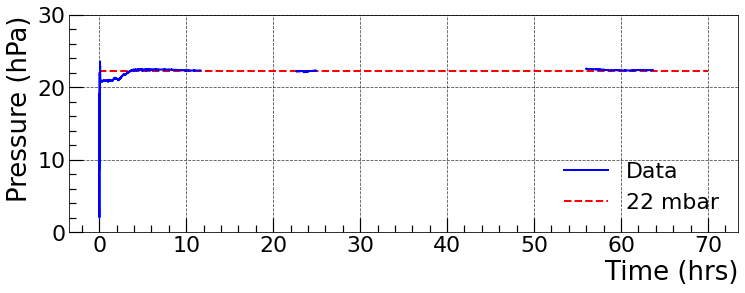}
\caption{The result of the gas leak test of the 3D printed frame conducted using Argon gas at 22~mbar overpressure. The blue line corresponds to the measured data and the red dotted line corresponds to the applied pressure.}
\label{fig:gasleak}
\end{figure}

\begin{figure}%
    \centering
    \subfloat[\centering]{{\includegraphics[width=2in]{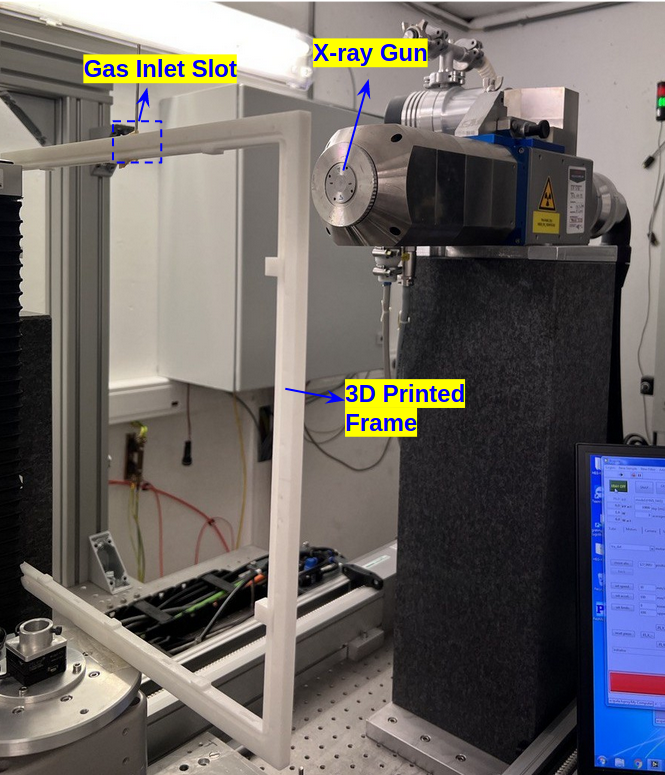}\label{fig:a1} }}%
    \qquad
    \subfloat[\centering]{{\includegraphics[width=0.8in]{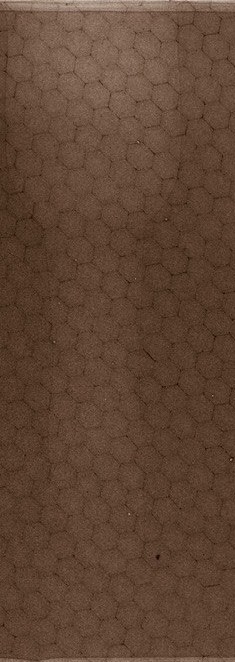}\label{fig:a2} }}%
    \caption{(a) The setup for the X-ray scanning of the 3D printed frame. The X-ray gun, detector frame and the location of the gas inlet in the frame are shown. (b) The image of the gas inlet obtained using X-ray scanning.}%
    \label{fig:xray}%
\end{figure}

In first instance, the mRPC prototype will be operated using a standard RPC gas mixture made of 95.2\% C$_2$H$_2$F$_4$, 4.5\% i-C$_4$H$_10$, and 0.3\% SF$_6$. However, the global warming potential of this mixture is around 1400, mainly because of the large R134A (C$_2$H$_2$F$_4$) component, and recent European regulations are increasingly restricting the usage or emission of greenhouse gases. While the search for more eco-friendly gas mixtures for RPCs is currently ongoing (see e.g.~\cite{ecogas}), we are opting for solutions that minimize the gas consumption. 
As such, a low-volume, leak-tight design is implemented, which already reduces gas consumption. The detector frame mentioned above is made of Polyamide powder and is fabricated as one single piece using laser-based 3D printing technology. The slots for the gas inlet and outlet and a gas channel that divides the gas input into four lines are also implemented in the frame to ensure uniform gas distribution inside the detector gas volume. For further leak protection, rubber O-rings and vacuum grease is used between the top and bottom PCB plates and the frame. Figure~\ref{fig:frame} shows the Computer-Aided Design (CAD) model of the frame. The channel at the top of the frame is for the O-rings and the square blocks on the four sides serve as spacers to hold the stack of glass plates in position.
To verify the gas tightness of the frame, a leak test was conducted using Argon gas at 22~mbar over-pressure. Figure~\ref{fig:gasleak} shows the result of the leak test. As can be seen, the chamber pressure remained constant without any drop for more than 60~hours, demonstrating the gas tightness of the design.
The frame was also scanned using X-rays to verify that the gas channel was implemented inside the frame as designed. Figure~\ref{fig:a1} shows the X-ray scanner and the scanning procedure of the frame. The X-ray gun was pointing towards the location of the gas inlet. The obtained X-ray image of the gas inlet is shown in Figure~\ref{fig:a2}. The gas inlet constructed inside the frame by the 3D printer was similar to what was designed in the CAD model. 

After the full assembly of the prototype, next steps will include chamber dark current and noise measurements, efficiency scans using cosmic muons and performance studies in terms of position and time resolution. Finally, the chamber will also be tested with more eco-friendly gas mixtures and with low gas flow or sealed mode, where the latter is of particular interest for muography applications.

\begin{figure}[htpb!]
\centering
\includegraphics[width=3in]{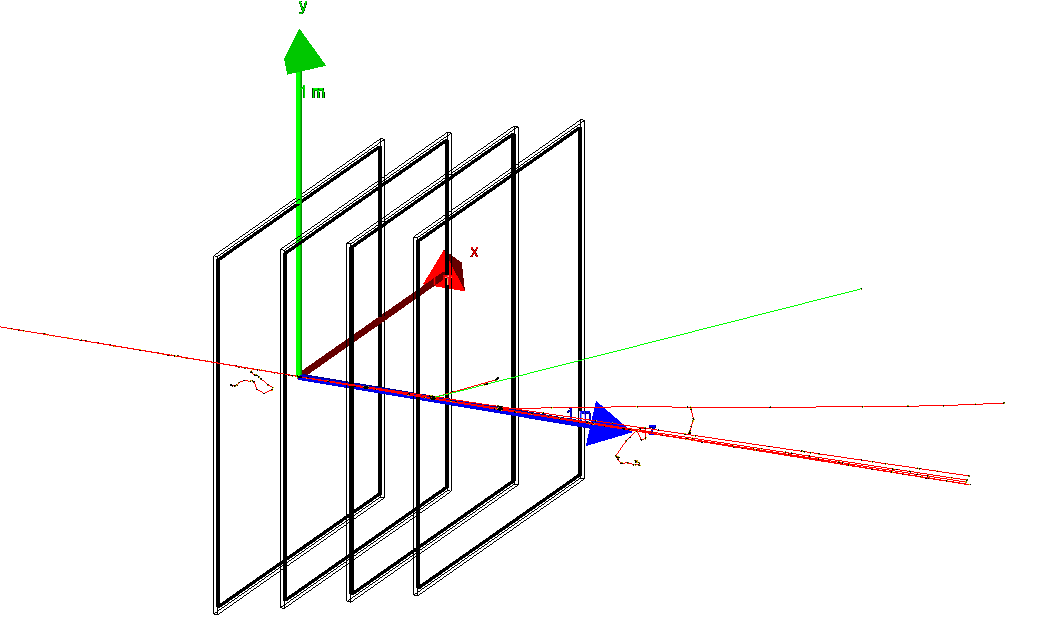}
\caption{Interaction of ten 800~MeV muons with the mRPC muon telescope in the Geant4 simulation. The red, green and blue axes in the figure represent the X, Y and Z axes in Geant4, respectively.}
\label{fig:geant4}
\end{figure}

 To optimise the
detector and telescope geometry and to study the role of time resolution can play in
background rejection, a Geant4 simulation of our 4-plane mRPC-based muon telescope is being developed. This simulation will eventually also be fine-tuned with actual telescope data and will also be needed in the effective muography image reconstruction once the setup is fully operational. Each mRPC station in the simulation setup is made up of a stack of seven glass plates of dimension $1\times 1$~m$^2$ separated by  gas gaps of 250~$\mu$m. The composition of the standard mixture is included in the simulation as well as all the remaining elements, such as Mylar foils, Copper plates, etc. Just for illustration, Figure~\ref{fig:geant4} shows the interaction of ten muons of energy 800~MeV in the simulated telescope. 

\begin{figure}[htpb!]
  \centering
  \subfloat[a][]{\includegraphics[width=3in]{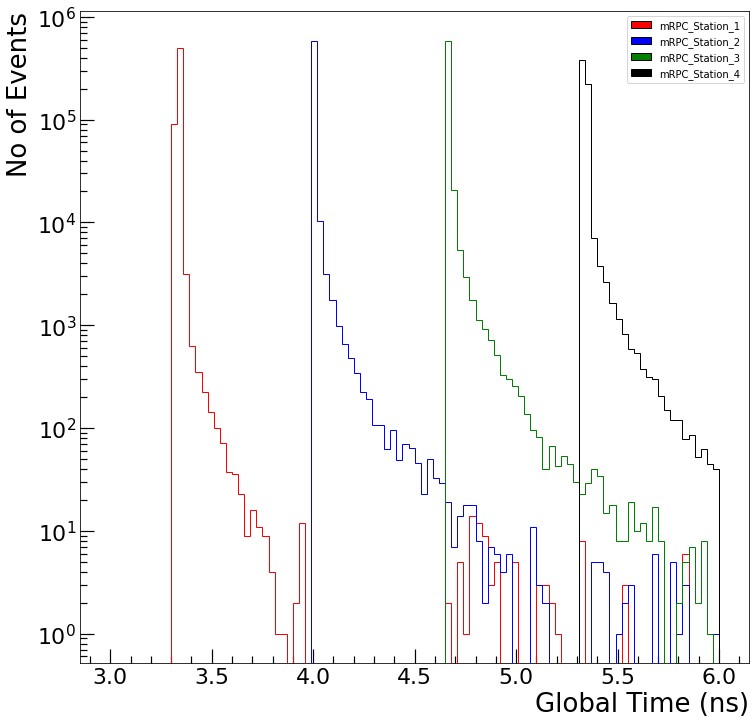} \label{fig:a2}} \\
  \subfloat[b][]{\includegraphics[width=3in]{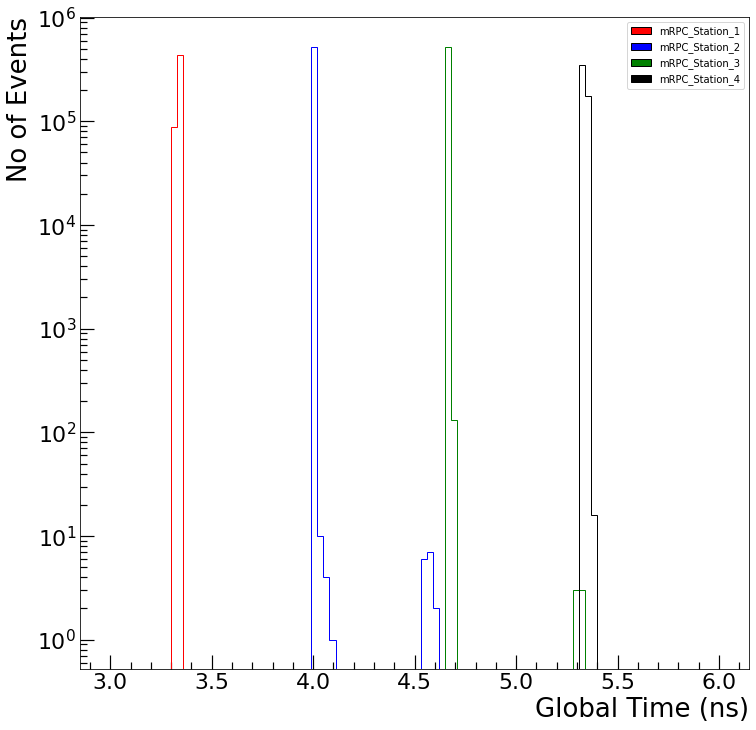} \label{fig:b2}}
  \caption{The global time distribution of Geant4 raw hits in four mRPC planes: (a) data without applying any energy thresholds, (b) data selected after applying an energy threshold of 100 MeV.
  The red, blue, green and black distributions correspond to stations 1, 2, 3 and 4, respectively.} \label{fig:simplots}
\end{figure}

In the telescope, the detector planes are placed vertically parallel to each other, at an inter-station distance of 20~cm, and with each subsequent station shifted upwards by 5~cm with respect to the previous one, in order for the telescope to slightly look upwards.

In very first preliminary study, 10$^5$ mono-energetic muons of 4~GeV, i.e. about the average cosmic-ray muon energy, were shot at the telescope in the direction perpendicular to the planes, and the detector responses were analysed. Since each plane has six gas gaps, multiple hits per plane are produced. Figure~\ref{fig:a2} shows the global time distribution associated with the hits generated in each plane.  A threshold of 100 MeV is applied for hit selection to eliminate the hits induced by the low-energy secondary particles. Figure~\ref{fig:b2} shows the global time distribution after applying this threshold. 

The results explain that a separation of 20~cm between the planes can already offer a time difference of $\sim 0.5$~ns. For further analysis, the particle generator software CRY will be used to generate cosmic muons. CRY provides cosmic-ray particle shower distributions at one of three elevations (sea level, 2100 m, and 11300 m) for input to the detector simulation~\cite{CRY}. To compare with the performance of the real mRPC detector, the Geant4 raw hit level data will be digitised. In the simulation, the possibility of adding the actual detector effects in terms of noise and factors affecting time resolution needs to be studied.
\section{Thick Gas Electron Multipliers}
Like the standard Gas Electron Multiplier, the Thick Gas Electron Multiplier (THGEM) belongs to the family of Micropattern Gaseous Detectors (MPGDs) using a hole-based gas avalanche multiplication~\cite{THGEM1}. Compared to standard RPCs, MPGDs offer improved spatial resolution down to order \SI{100}{\micro\metre} and can be operated at lower, moderate high voltages with simpler gas mixtures or even mono-gases, which makes them particularly interesting for muography applications (see e.g.,~\cite{muongraphyapplication}). Moreover, the simplified THGEM manufacturing procedure with mechanical hole drilling and Copper etching on Printed Circuit Boards makes them a more economical alternative for standard GEMs for which the metallic amplification foils are produced via a complex chemical etching procedure.

\begin{figure}[htpb!]
\centering
\includegraphics[width=3.5in]{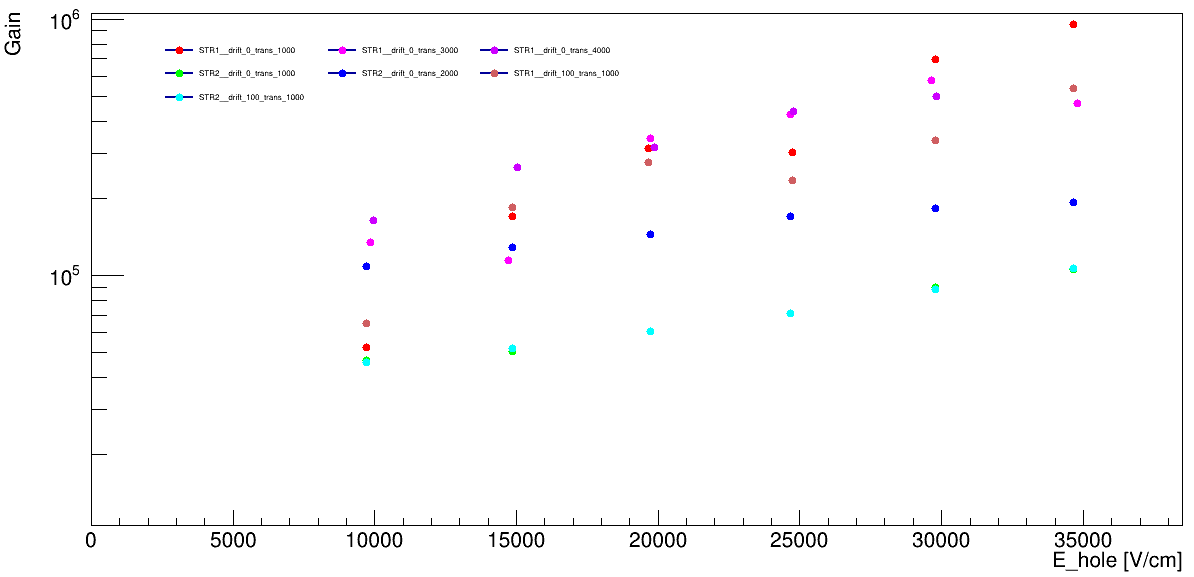}
\caption{Simulated effective gain of a single-layer THGEM as a function of the THGEM multiplication field, compared for different drift fields ("$drift$") and induction fields ("$trans$"). The simulation has been done for a THGEM PCB thickness of 0.5~mm, drift gap 3~mm, induction gap 1~mm, GEM hole rim 0.1~mm, and for an $Ar/CO_2$ (70/30) gas mixture.}
\label{THGEMgain}
\end{figure}

In order to study and optimize the basic layout and performance of the THGEM in terms of e.g. hole configurations, gap sizes, field values, gas mixture, and spatial resolution, a Garfield++-based~\cite{garfield} simulation study was initiated.
In a first step, the effective gain the detector was studied for different configurations. The chamber was simulated with a drift gap of 3~mm, induction gap of 1~mm, GEM hole rim size of 0.1~mm and an $Ar/CO_2$ (70/30) gas mixture.
Figure \ref{THGEMgain} shows the effective gain of a single-layer THGEM as a function of the multiplication field $E_{hole}$. Here, the effective gain is defined as the ratio between the number of electrons reaching the readout electrode and the number of primary electrons produced in the drift region, and $E_{hole}$ refers to the value of the electric field applied to the THGEM PCB. Two types of THGEM PCB structures are compared in the study. The first structure (STR1) has a ratio between hole diameter ($d$) and PCB thickness ($t$) of $r_{STR1} \sim 1$ ($d=0.4$~mm, $t=0.5$~mm), while the second structure (STR2) has $r_{STR2} = 0.5$ ($d=1$~mm, $t=0.5$~mm). The colour coding indicates different drift fields (labeled $drift$) and induction fields (labeled $trans$), where the drift field indicates the electric field between the cathode board and the THGEM PCB, while the induction field represents the field between the THGEM PCB and the readout board. 
As shown in Figure~\ref{THGEMgain}, the effective gain values of STR1 are generally higher compared to STR2, indicating that the effective gain of the THGEM is higher when the ratio ($r$) between hole diameter and PCB thickness $\sim 1$. 

Studies of the drift velocity for different gas mixtures and of the chamber gain as function of the induction and THGEM field are currently ongoing.

\begin{figure}[htpb!]
\centering
\includegraphics[width=2.3in]{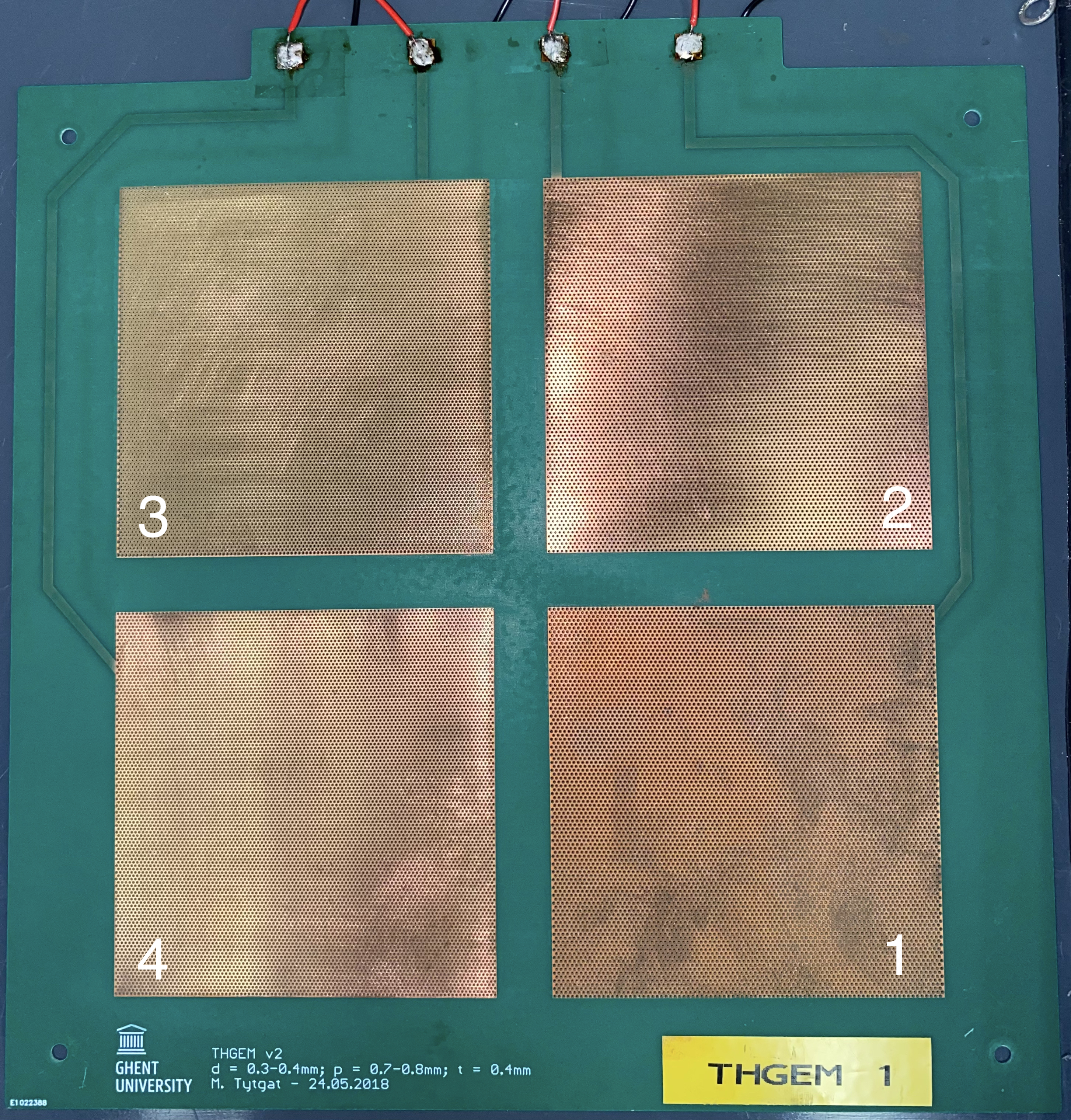}
\caption{Multiple THGEMs embedded into one PCB for prototyping studies. A total of four $7\times 7$~cm$^2$ THGEMs with different hole diameters and pitch distances are fabricated for a comparative performance study.}
\label{GEMfoil}
\end{figure}

In parallel to the simulation, small-size detector prototypes are also under development using commercially produced THGEM PCBs. Figure~\ref{GEMfoil} shows a first manufactured PCB with the THGEM foils embedded. A total of four $7\times 7$~cm$^2$ THGEMs of thickness $t = 0.4$~mm are embedded, with different hole diameters ($d$) and pitch distances ($p$) and with a hole rim size of 0.1~mm. The specifications of the THGEM foils are summarised in Table~\ref{tab:foilspec}.

\begin{table}[h]
    \centering
    \begin{tabular}{c|cc}\hline
        THGEM & hole diameter($d$) [mm] &  pitch distance ($p$) [mm] \\ \hline
         1 & 0.3 & 0.7 \\
         2 & 0.4 & 0.7 \\
         3 & 0.3 & 0.8\\
         4 & 0.4 & 0.8\\\hline
    \end{tabular}
    \caption{Specifications of the prototype THGEMs.}
    \label{tab:foilspec}
\end{table}

\begin{figure}[htpb!]
\centering
\includegraphics[width=3.2 in]{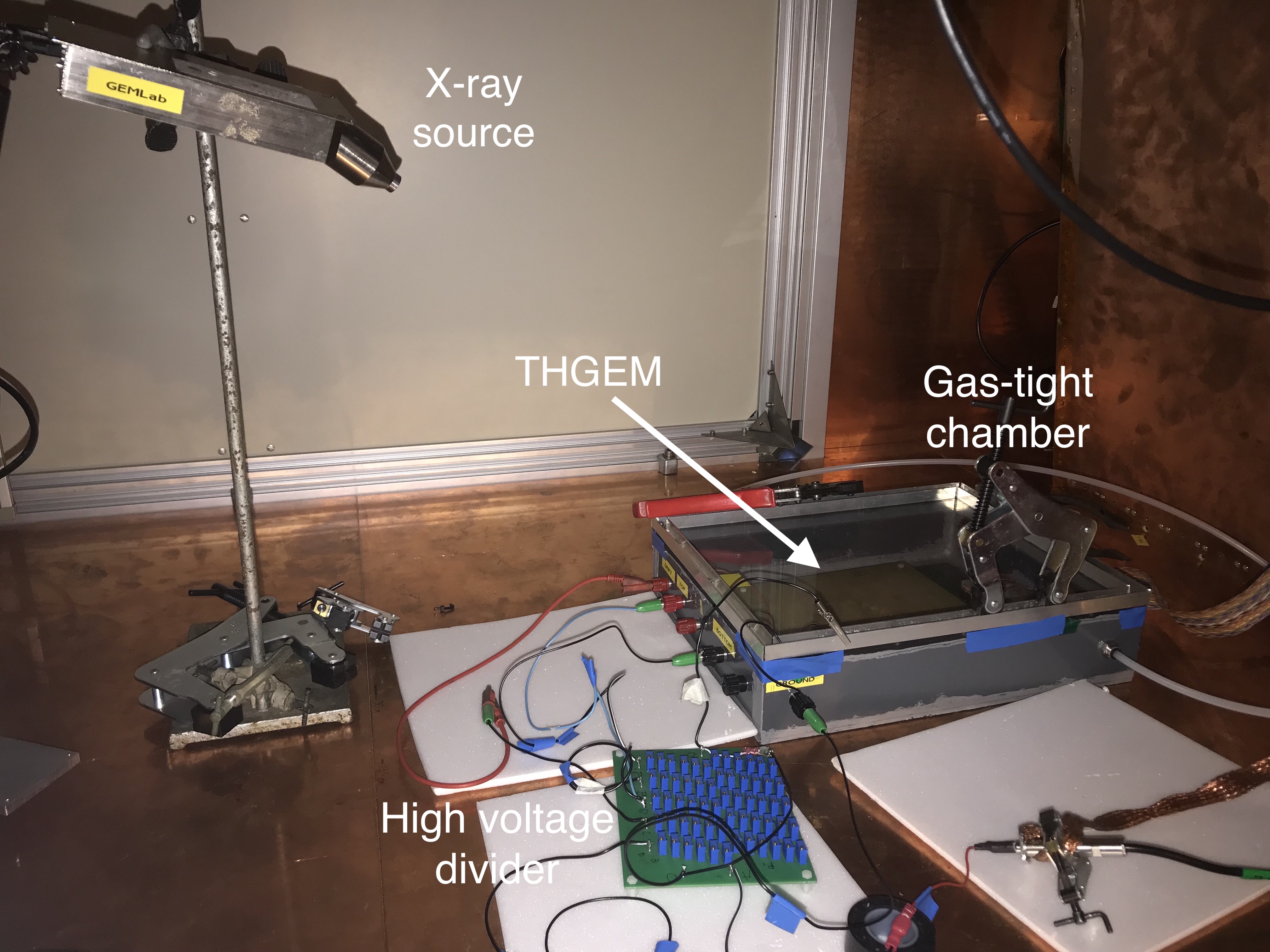}
\caption{Experimental setup of the prototype THGEM inside a large Copper test cage. The THGEM detector assembly is placed inside a gas-tight container. The AMPTEK X-ray gun is aimed directly at the THGEM chamber.}
\label{THGEMsetup}
\end{figure}

\begin{figure}[htpb!]
\centering
\includegraphics[width=2.3in]{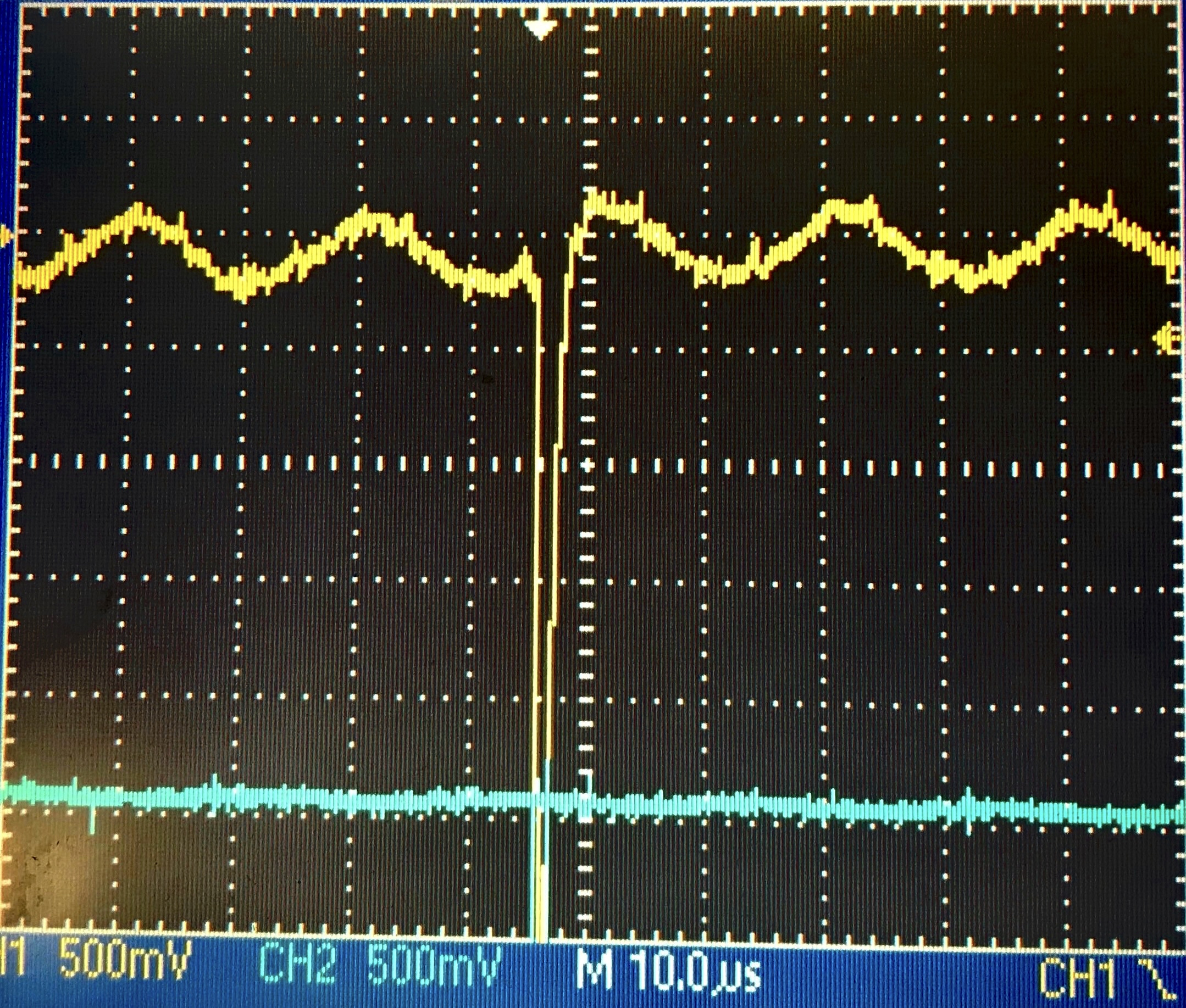}
\caption{THGEM chamber signal observed with an oscilloscope during irradiation with the X-ray source.}
\label{THGEMsignal}
\end{figure}

Initial performance studies of the THGEMs were conducted using an AMPTEK Mini-X X-ray tube~\cite{amptek}, in which the detector stability, efficiency, effective gain and possible discharge or charging-up effects were investigated. Figure~\ref{THGEMsetup} shows the experimental setup for the preliminary tests of a single-layer THGEM. The chamber was operated with a standard gas mixture $Ar/CO_2$ (70/30). A external resistor high-voltage divider is used to power and adjust the various THGEM fields.
First, a series of stability tests were performed to verify that the THGEMs could sustain an increasingly applied high voltage, i.e. that no shorts in the THGEMs were present. During the subsequent studies, the high voltage configuration as given in Table~\ref{tab:E_field_gain 2} was applied. Operating the chamber without the X-ray source, a background rate of 2.7~Hz was observed, while with the X-ray source on the rate increased to 12.5~Hz. To confirm that this increment was effectively due to the source, additional measurements were performed where $E_{THGEM}$ was set to zero, and where indeed no increase in the chamber rate was observed with or without the X-ray source. Figure~\ref{THGEMsignal} illustrates the signal from the X-ray source as observed on an oscilloscope.

\begin{table}[h]
    \centering
    \begin{tabular}{c|ccc}\hline
         & Gap [$mm$] &  Voltage [$kV$] & Field [$kV/cm$]\\ \hline
         $E_{drift}$& 6 & 1.7 & 2.83  \\
         $E_{THGEM}$& 0.4 &1.7 & 42.5 \\
         $E_{induction}$& 1 & 0.43 & 4.31\\\hline
    \end{tabular}
    \caption{High voltage configuration of the THGEM chamber during the initial performance study.}
    \label{tab:E_field_gain 2}
\end{table}

\section{Summary and Outlook}
Muon radiography is an imaging technique relying on freely available cosmic-ray muons to probe the internal structure of microscopically large objects. 2D or 3D density profiles of a target can be obtained via measurements of the transmitted muon flux through the object. Muography telescopes are traditionally based on either nuclear emulsions, scintillators or gaseous detectors. To benefit from the particular advantages that certain detector technologies offer, a muon radiography telescope is being developed using a combination of scintillators and advanced gaseous detectors, which should result in a highly performing system in terms of both spatial and temporal resolution. In addition, the system will be optimized in terms of gas consumption to facilitate its usage in remote muography campaigns. 

A baseline solution using scintillator bars readout via wavelength shifting fibers connected to MPPCs is currently under construction. An initial prototype has been developed with four MPPCs to study the scintillator and signal properties. The full detector will in a first step be studied using a PETIROC2A ASIC based DAQ system from CAEN.

High-time resolution Multi-gap Resistive Plate Chambers are considered to improve detector background rejection using timing. A double 6-gap prototype chamber with float glass electrodes is currently under construction. 
The glass electrodes are coated with a two-component resistive paint mixture using a manual spraying technique, which yields uniform and stable in time values for the surface resistivity of the electrodes. The thickness of the resistive coating was measured using a SEM method, which confirmed the layer uniformity and thickness of $\sim 35 \ \mu$m and validated our coating procedure. To reduce its gas consumption and improve the portability of the detector, a 3D powder-printed detector frame was developed to enclose the glass plate stack and form the chamber gas volume. The quality of the 3D printing technique has been verified using X-rays. Next to the hardware construction, a Geant4-based simulation is being developed for a four-plane mRPC muon telescope. This simulation will be used to study the usage of time resolution for background rejection and to 
optimise the layout of our telescope. 

In addition to mRPCs, high gain Thick-GEMs are considered as an alternative to our baseline solution to exploit their superior spatial resolution. Such detectors can be also be operated with simple, ecofriendly gas mixtures or even mono-gases, and at lower voltages compared to RPCs, which makes them particularly attractive for muography applications. 
Garfield++ simulation studies are ongoing to optimize the detector geometry, GEM hole configurations and field values. The simulation studies are also performed for different detector gas mixtures. Initial detector studies with small THGEM prototypes have been started. 



Once the basic design of the three types of muon detectors has been established and operational prototypes have been constructed, the configuration and performance of an advanced muon telescope made of a combination of stations of different technologies can be investigated. 


\end{document}